\documentclass[10pt,showpacs,pra,twocolumn]{revtex4-1}

\usepackage{amsfonts}
\usepackage{amsmath}
\usepackage[english]{babel}
\usepackage[latin1]{inputenc}
\usepackage[dvips]{graphics,color}
\usepackage{graphicx}
\usepackage{amssymb}

\def\be{\begin{equation}}
\def\ee{\end{equation}}
\def\bea{\begin{eqnarray}}
\def\eea{\end{eqnarray}}
\newcommand{\beqn}{\begin{eqnarray}}
\newcommand{\eeqn}{\end{eqnarray}}
\newcommand{\beqnn}{\begin{eqnarray*}}
\newcommand{\eeqnn}{\end{eqnarray*}}

\begin{document}

\title{Influence of the field-detector coupling strength on the dynamical Casimir effect} 

\author{A. S. M. de Castro}
\email{asmcastro@uepg.br}
\affiliation{Universidade Estadual de Ponta Grossa, Departamento de F\'{\i}sica, CEP
84030-900, Ponta Grossa, PR, Brazil.}
\author{A. Cacheffo}
\email{cacheffo@pontal.ufu.br}
\affiliation{Universidade Federal de Uberl\^andia, Faculdade de Ci\^encias Integradas do
Pontal, CEP 38302-000, Ituiutaba, MG - Brazil.}
\author{V. V. Dodonov}
\email{vdodonov@fis.unb.br}
\affiliation{Universidade de Bras\'{\i}lia, Instituto de F\'{\i}sica, Caixa Postal 04455, CEP
70910-900 Bras\'{\i}lia, DF, Brazil.}

\begin{abstract}
We consider the problem of photon creation from vacuum inside an ideal cavity with vibrating walls in the
resonance case, taking into account the interaction between the resonant field mode and a detector modeled by a quantum
harmonic oscillator. The frequency of wall vibrations is taken to be twice the cavity normal frequency, modified due to
the coupling with the detector. The dynamical equations are solved with the aid of the multiple scales method.
Analytical expressions are obtained for the photon mean numbers and their variances for the field and detector modes, which
are supposed to be initially in the vacuum quantum states. We analyze different regimes of excitation, depending
on the ratio of the modulation depth of the time-dependent cavity eigenfrequency to the
coupling strength between the cavity mode and detector. We show that statistical properties of the detector quantum state
(variances of the photon numbers, photon distribution function, and the degree of quadrature squeezing) can be quite
different from that of the field mode. Besides, the mean number of quanta in the detector mode increases with some time delay,
compared with the field mode.

\end{abstract}

\maketitle

\section{Introduction}
\label{sec-int}

The effect of photon creation from vacuum in a cavity with rapidly varying geometrical or material properties,
called nowadays as Dynamical Casimir Effect (DCE), attracted attention of many researchers for a long time:
see recent reviews \cite{VD-PS10,DAR-11,PNA-12}. The first experimental results for the open strip-line waveguide
with time-dependent boundary conditions, simulating a single oscillating ideal mirror in the one-dimensional
space, were reported recently in \cite{CMW-N10}. It is quite probable that new experiments in other geometries will be done
soon, too. Therefore, the problem of the back action of different detectors on the rate of photon generation
becomes actual. The very first studies were performed in \cite{D95,DK96} under the condition that the field-detector
coupling is much stronger than the effective coupling between the selected resonance field mode in the cavity
and moving boundary (due to strong limitations on the attainable amplitude of surface oscillations in real materials).

However, since that time there were proposed several schemes on simulating the DCE by means of changing not the
positions of boundaries but their material properties \cite{CMW-N10,Padua05}. In such cases the effective velocity
of boundaries can be increased by several orders of magnitude, therefore the situations where the modulation
depth of the instantaneous cavity eigenfrequency is bigger than the normalized field-detector coupling
coefficient can be considered now quite realistic, as well.

There are two simple models of detectors. One of them describes the detector as a two-level ``atom'' \cite{D95}.
It can be applied for the experimental setup proposed in \cite{Onof06}. This case was extensively studied in different
regimes in \cite{shake,Kawa11,Roberto,AD01}, and generalizations to the three- and multi-level ``atoms'' were considered in
\cite{AD02}. 

Here we consider another simple model, where the detector is represented by a harmonic oscillator tuned in resonance with
the selected field mode \cite{D95,DK96}. This model seems to be adequate to the case of the
so called MIR experiment \cite{Padua05,MIR2}, where the microwave quanta created via the
DCE are supposed to be detected by means of a small antenna put inside the cavity.
Since the inductive antenna (a wire loop) used in that experiment is a part of a LC-contour,
it can be reasonably approximated as a harmonic oscillator.  Therefore we study the quantum system described by the Hamiltonian
(in dimensionless units; in particular, we assume $\hbar=1$)
\begin{equation}
\hat{H}=\frac{1}{2}[\hat{p}_{1}^{2}+\omega^{2}(t)\hat{x}_{1}^{2}+\hat{p}_{2}^{2}+\omega_{0}^{2}\hat{x}_{2}^{2}-4\omega_{0}\kappa\hat{x}_{1}\hat{p}_{2}].  \label{Htot}
\end{equation}
Here $\hat{x}_1$ and $\hat{p}_1$ are the quadrature component operators of the selected resonance field mode, whereas
$\hat{x}_2$ and $\hat{p}_2$ are the quadrature component operators of the detector (antenna). The interaction part of the
Hamiltonian is assumed to be proportional to the product $\hat{x}_1\hat{p}_2$ \cite{D95,DK96}
in view of the standard minimal coupling term $-(e/c){\bf p A}$, assuming that operator $\hat{x}_1$ is proportional
to the vector potential of the field mode.
The instantaneous cavity eigenfrequency $\omega(t)$ is chosen in the form
\begin{equation}
\omega(t)=\omega_{0}[1+2\gamma\cos(\Omega t)],
\label{parfreq}
\end{equation}
where $\omega_0$ is the unperturbed mode frequency and $\Omega$ is the modulation frequency.
The real modulation depth $\gamma$ is assumed to be small, as well as the real coupling coefficient $\kappa$.
The influence of all other cavity modes can be neglected if the spectrum of cavity eigenfrequencies is not
equidistant \cite{DD-PLA12}.

If $\gamma=0$, then one can diagonalize Hamiltonian (\ref{Htot}), introducing two normal modes, whose frequencies
are given by the exact formula
$
\omega_{\pm} = \omega_0\sqrt{1 \pm 2\kappa}$,
so that for $|\kappa|\ll1$ we have
\begin{equation}
\omega_{\pm} \approx \omega_{0}(1 \pm\kappa).  \label{gf}
\end{equation}
We assume that the modulation frequency $\Omega$ is twice bigger than one of the two splitted normal frequencies.
We choose $\Omega=2\omega_{-}$ and normalize the frequencies and dynamical variables
in such a way that $\omega_0=1$. Our goal is to study the influence of the dimensionless ratio
$\beta=\gamma/(2\kappa)$ on the number of excitations (``photons'') in the field mode
$\mathfrak{\bar{n}}_{1} = \frac12\langle \hat{x}_1^2 + \hat{p}_1^2\rangle -\frac12$
and in the detector
$\mathfrak{\bar{n}}_{2} = \frac12\langle \hat{x}_2^2 + \hat{p}_2^2\rangle -\frac12$
for arbitrary (small or big) values of this parameter.

The paper is organized as follows. In Sec. \ref{sec-inv} we
derive the main dynamical equations and obtain their approximate analytical solutions
 in the weak coupling and weak modulation regimes, comparing these solutions with results of numerical calculations.
 In Sec. \ref{sec-pgv} we use these solutions to calculate the time-dependent photon mean numbers for the initial
 vacuum state in both the cavity and detector modes. In Sec. \ref{sec-fluc} we calculate the variances
 of the photon number operator, the photon distribution functions and the degree of squeezing in each mode.
The results are discussed in Sec. \ref{sec-con}.

\section{Main dynamical equations and their approximate solutions}
\label{sec-inv}

Hamiltonian (\ref{Htot}) is a special case of generic quadratic Hamiltonians (see Ref. \cite{QIBook} for a detailed treatment), which can
be written as 
\begin{equation}
\hat{H}=\frac{1}{2}\hat{\mathbf{q}}\mathcal{B}\hat{\mathbf{q}}, \quad \hat{\mathbf{q}} \equiv(\hat{\mathbf{p}},\hat{\mathbf{x}})=(\hat{p}_{1},\hat{p}_{2},\hat{x}_{1},\hat{x}_{2}).
\label{HqBq}
\end{equation}
Here $\mathcal{B}(t)$ is a symmetrical $4\times4$ matrix, which can be splitted in four $2\times2$ blocks
as
\begin{equation*}
\mathcal{B}=\left \Vert
\begin{array}{cc}
{b}_{1} & {b}_{2} \\
{b}_{3} & {b}_{4}
\end{array}
\right \Vert , \qquad {b}_{1}=\widetilde{{b}}_{1}, \; {b}_{4}=\widetilde{{b}}_{4}, \; {\ b}_{2}=\widetilde{{b}}_{3},
\end{equation*}
where the tilde (\symbol{126}) means matrix transposition.
In the case involved we have
\begin{gather}
b_{1}=\left\Vert
\begin{array}{cc}
1 & 0 \\
0 & 1
\end{array}
\right\Vert ,\quad b_{2}=\left\Vert
\begin{array}{cc}
0 & 0 \\
-2\kappa & 0
\end{array}
\right\Vert ,
\quad b_{4}=\left\Vert
\begin{array}{cc}
\omega^{2}(t) & 0 \\
0 & 1
\end{array}
\right\Vert .
\nonumber
\end{gather}

For any quadratic homogeneous Hamiltonian, the Heisenberg operators $\hat{\mathbf{q}}(t)$
are related to the initial operators $\hat{\mathbf{q}}(0)$ by means of a linear transformation:
\begin{equation}
\hat{\mathbf{q}}(t) = \mathcal{L}(t)\hat{\mathbf{q}}(0), \qquad
\hat{q}_{\mu}(t)=\mathcal{L}_{\mu\alpha}(t)\hat{q}_{\alpha}(0)
\label{solHeis}
\end{equation}
(the summation over repeated Greek indices is implied),
where coefficients $\mathcal{L}_{\mu\alpha}(t)$ form the time-dependent symplectic matrix $\mathcal{L}(t) \equiv \Vert \mathcal{L}_{\mu\alpha}(t)\Vert$.
It is more convenient, however, to use the
inverse matrix $\Lambda =\mathcal{L}^{-1} $. Then  Eq. (\ref{solHeis}) can be rewritten as
\begin{equation}
\hat{\mathbf{q}}_{0}(t)=\Lambda(t)\hat{\mathbf{q}},  \label{int-mot}
\end{equation}
showing that $\hat{\mathbf{q}}_{0}(t)$ is the operator integral of motion in the Schr\"odinger picture.
Matrix $\Lambda(t)$ satisfies the following differential matrix equation and initial condition:
\begin{equation}
\dot{\Lambda}=\Lambda\Sigma\mathcal{B},\quad\Sigma=\left\Vert
\begin{array}{cc}
0 & I_{2} \\
-I_{2} & 0
\end{array}
\right\Vert ,\quad\Lambda(0)=I_{4},
\label{qLam}
\end{equation}
where $I_{N}$ means the $N\times N$ identity matrix\ and the elements of matrix $\Sigma\ $satisfy the commutation relation $\Sigma_{ij}=i[\hat{x}_{i},\hat {p}_{j}]=-\delta_{ij}$. Moreover, the following symplectic identities are
immediate consequencies of Eq. (\ref{qLam}):
\begin{equation}
{\Lambda}\Sigma\widetilde{\Lambda}\equiv\Sigma,\quad{\Lambda}^{-1}\equiv \Sigma\widetilde{\Lambda}^{-1}=-\Sigma\widetilde{\Lambda}\Sigma.
\label{ident}
\end{equation}
Writing matrices $\Lambda(t)$ and $\mathcal{L}(t)$ in the block form,
\begin{equation}
\Lambda(t)=\left\Vert
\begin{array}{cc}
\lambda_{1} & \lambda_{2} \\
\lambda_{3} & \lambda_{4}
\end{array}
\right\Vert , \qquad
\mathcal{L}(t)=\left\Vert
\begin{array}{cc}
\widetilde\lambda_{4} & -\widetilde\lambda_{2} \\
-\widetilde\lambda_{3} & \widetilde\lambda_{1}
\end{array}
\right\Vert ,
\label{LamL}
\end{equation}
we obtain from Eqs. (\ref{qLam}) the equations (for $b_1=I_2$)
\be
d\lambda_1/dt = \lambda_1 b_3 - \lambda_2, \qquad
d\lambda_2/dt = \lambda_1 b_4 - \lambda_2 b_2,
\label{eq-lam12}
\ee
and similar equations for the blocks $\lambda_3$ and $\lambda_4$.
Excluding matrices $\lambda_2$ and $\lambda_4$ we arrive at identical second-order equations for matrices
$\lambda_1$ and $\lambda_3$:
\be
\frac{d^{2}\lambda_{1,3}}{dt^{2}}-\frac{d\lambda_{1,3}}{dt}\mathbf{R}_{1}+\lambda_{1,3}\mathbf{R}_{3}  =0,
\label{E1}
\ee
where
\be
\mathbf{R}_{1} = b_3-b_2 =
2\kappa\left\Vert
\begin{array}{cc}
0 & -1 \\
1 & 0
\end{array}
\right\Vert ,
\label{N1}
\ee
\be
\mathbf{R}_{3}  = b_4 -b_3 b_2
=\left\Vert
\begin{array}{cc}
\omega^{2}(t) -(2\kappa)^2 & 0 \\
0 & 1
\end{array}
\right\Vert .
\label{N3}
\ee
The difference between $\lambda_1(t)$ and $\lambda_3(t)$ is in the initial conditions:
\begin{align*}
\lambda_{1}(0) & =I_{2},\quad\dot{\lambda}_{1}(0)=b_{3}, \\
\lambda_{3}(0) & =0,\quad\ \dot{\lambda}_{3}(0)=-I_{2}.
\end{align*}
We suppose that $|\gamma|, |\kappa| \ll 1$.
In such a case, Eq. (\ref{E1}) can be solved with a sufficient accuracy analytically with the aid of the
method of multiple scales \cite{nayfeh,Janow03}. For this purpose we introduce a formal small parameter
$\varepsilon$, writing $\gamma=\varepsilon\gamma_0$, $\kappa=\varepsilon\kappa_0$ [with
$\gamma_0,\kappa_0 \sim \mathcal{O}(1)$] and assuming that the solutions depend on the set of
{\em scaled times}, $T_0=\omega_0 t$, $T_1=\varepsilon\omega_0 t$, $T_2=\varepsilon^2\omega_0 t, \ldots$,
which can be considered as {\em independent variables}. This means that the time derivatives
$d/dt$ and $d^2/dt^2$  can be expressed as follows:
\begin{align}
\frac{d}{dt}& =\frac{\partial }{\partial T_{0}}+\varepsilon \frac{\partial }{\partial T_{1}}
+\mathcal{O}(\varepsilon ^{2}),  \notag \\
\frac{d^{2}}{dt^{2}}& =\frac{\partial ^{2}}{\partial T_{0}^{2}}
+2\varepsilon \frac{\partial ^{2}}{\partial T_{1}\partial T_{0}} +\mathcal{O}(\varepsilon ^{2}).  \notag
\end{align}
Also, we assume that matrices  $\lambda _{k}\left( t\right)$ can be written in the form
\begin{equation}
\lambda _{k}(t)=\lambda _{k,0}(T_{0},T_{1},...)+\varepsilon \lambda _{k,1}(T_{0},T_{1},...)+\mathcal{O}(\varepsilon ^{2}).
\end{equation}
For small enough values of parameter $\varepsilon$ it is sufficient to calculate the first term of
the above expansion, taking into account the dependence on times $T_0$ and $T_1$ only.
Therefore we neglect the terms of the second order with respect to $\varepsilon$
in Eq. (\ref{E1}), replacing exact matrix (\ref{N3}) by the approximate form
\be
\mathbf{R}_{3}  \approx \left\Vert
\begin{array}{cc}
1+4\gamma_0\varepsilon\cos(2T_0 - 2\kappa_0 T_1) & 0 \\
0 & 1
\end{array}
\right\Vert .
\label{N3a}
\ee
Omitting the details of cumbersome calculations, we bring here
the general structure of solutions only. It appears that
the elements $\lambda _{k}^{(ij)}$ of each matrix $\lambda_k(t)$ (where $k=1,2,3,4$
and $i,j=1,2$) can be written 
as (remember that we put $\omega_0=1$)
\begin{eqnarray}
\lambda_k^{(ij)} &=& \frac1{2\eta} \cos(\omega_{-}t)\Big\{F_k^{(ij)} \cosh(\tau_{\mu}) + G_k^{(ij)} \sinh(\tau_{\mu})/\mu
\nonumber \\ &&
+ U_k^{(ij)} \cos(\tau_{\nu}) + V_k^{(ij)} \sin(\tau_{\nu})/\nu\Big\}
\nonumber \\
&+& \frac1{2\eta} \sin(\omega_{-}t)\Big\{f_k^{(ij)} \cosh(\tau_{\mu}) + g_k^{(ij)} \sinh(\tau_{\mu})/\mu
\nonumber \\ &&
+ u_k^{(ij)} \cos(\tau_{\nu}) + v_k^{(ij)} \sin(\tau_{\nu})/\nu\Big\}.
\label{lambdas}
\end{eqnarray}
The meaning of symbols is as follows.
The ``fast time'' $t$ appears in two oscillating functions $\cos(\omega_{-}t)$ and $\sin(\omega_{-}t)$ only.
The amplitudes of these fast oscillations are modulated by combinations of hyperbolic and trigonometric functions, which
depend on the ``dimensionless slow time'' $\tau=\kappa t$ (which is proportional to $T_1$). Namely, hyperbolic functions
depend on variable $\tau_{\mu}=\mu\tau$, whereas trigonometric functions depend on
$\tau_{\nu}=\nu\tau$. In turn, coefficients $\eta$, $\mu$ and $\nu$ depend on the ratio $\beta=\gamma/(2\kappa)$:
\begin{eqnarray}
\mu(\beta) &=& \sqrt{2\eta+2(\beta^{2}-1)},
\\
\nu(\beta) &=& \sqrt{2\eta-2(\beta^{2}-1)},
\\
\eta(\beta) &=&\sqrt{\beta^{2}(\beta^{2}-1)+1}.
\end{eqnarray}
 The factors $\mu^{2}$ and $\nu^{2}$ are both positive for any values of $\beta$ and satisfy the identity $\mu \nu=2\beta$.
All other constant coefficients in Eq. (\ref{lambdas}), such as $F_k^{(ij)}$, $f_k^{(ij)}$, and so on,
also depend on the ratio $\beta$. Explicit forms of coefficients with $k=1$ and $k=3$ are given in Appendix.
Since matrix $b_3$ is of the order of $\varepsilon$, we can write, in view of Eq. (\ref{eq-lam12}),
approximate formulas $\lambda_{2,4} \approx -d\lambda_{1,3}/dt$. Consequently, within the chosen accuracy (i.e.,
neglecting terms proportional to $\varepsilon$ in all the amplitude coefficients) the elements of matrices
$\lambda_{2,4}$ can be obtained from the respective formulas (\ref{lambdas}) for $\lambda_{1,3}$ by means of replacements
$\cos(\omega_{-}t) \to \sin(\omega_{-}t)$ and $\sin(\omega_{-}t) \to -\cos(\omega_{-}t)$.

\begin{figure}[htb]
\par
\begin{center}
\includegraphics[height=3.32truein,width=3.32truein,angle=0]{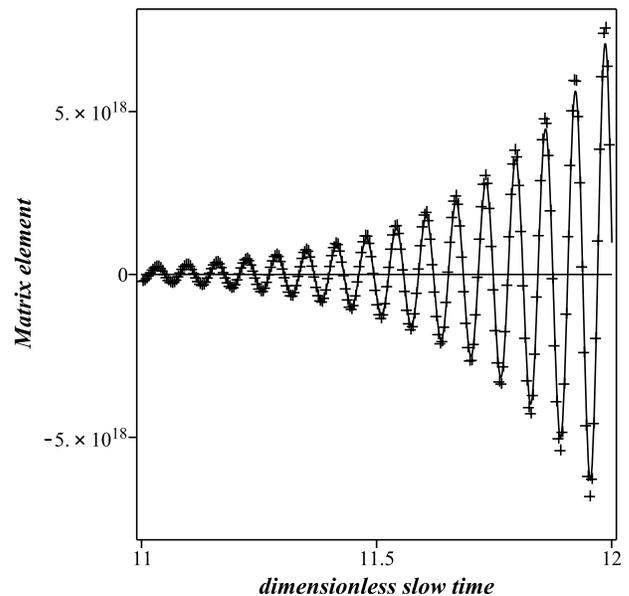}
\end{center}
\par
\vspace{0.1cm}
\caption{Numerical (points) and analytical (solid line) solutions for the matrix element $\lambda_1^{(11)}(t,\tau)$ in the case of $\gamma_0= 4.0$, $\kappa_0= 1.0 $ with $\beta =2 $ and $\varepsilon=0.01$.
 }
\label{fig-anal-numer}
\end{figure}

We have checked the accuracy of approximate analytical solutions (\ref{lambdas}), comparing them with exact numerical solutions
of Eq. (\ref{E1}). It appears that the coincidence is quite good for $\varepsilon < 10^{-2}$. An example of comparing
analytical and numerical solutions is given in Fig. \ref{fig-anal-numer}.

\section{The photon generation from vacuum state}
\label{sec-pgv}

\begin{figure}[htb]
\par
\begin{center}
\includegraphics[height=3.4truein,width=3.4truein,angle=0]{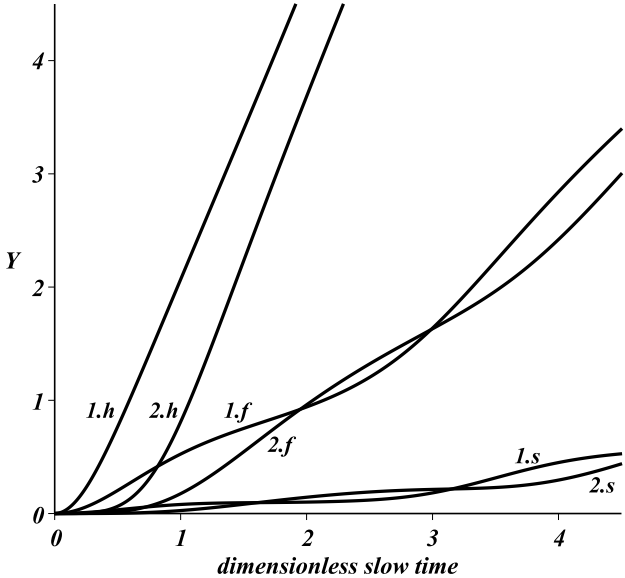}
\end{center}
\par
\vspace{0.1cm}
\caption{The behavior of function $Y(\tau)=\ln[1+\mathfrak{\bar{n}}_{m}(\tau)]$ for the cavity (1) and detector (2) with $\beta=0.2$ (\textit{s}), $\beta=0.5$ (\textit{f}) and $\beta = 1.0$ (\textit{h}). Note the time delay in increasing the number of quanta in the detector for $\beta=0.5$ and $\beta=1.0$.}
\label{lnmpn}
\end{figure}

Knowing matrices $\Lambda(t)$ or $\mathcal{L}(t)$ (\ref{LamL}) one can calculate immediately the time evolution of the
symmetric covariance matrix $\mathcal{M}=\Vert \mathcal{M}_{\mu \nu }\Vert$, whose elements
%
$\mathcal{M}_{\mu \nu }=\frac{1}{2}\left\langle \widehat{\mathrm{q}}_{\mu }\widehat{\mathrm{q}}_{\nu }+\widehat{\mathrm{q}}_{\nu }\widehat{\mathrm{q}}_{\mu }\right\rangle -\left\langle \widehat{\mathrm{q}}_{\mu }\right\rangle\left\langle \widehat{\mathrm{q}}_{\nu }\right\rangle $
 are the central second-order statistical moments. Namely,
\begin{equation}
\mathcal{M}(t)=\mathcal{L}(t)\mathcal{M}(0)\widetilde{\mathcal{L}}(t).
\label{covte}
\end{equation}
For the initial vacuum states of both modes matrix $\mathcal{M}(0)$ is proportional to the unity matrix:
$\mathcal{M}(0)=\frac12 I_4$.
Consequently, in view of Eqs (\ref{LamL}) and (\ref{covte}),
the mean number of quanta in the $m$th mode ($m=1,2$) can be expressed
in terms of diagonal elements of the sum of products of  matrices $\widetilde{\lambda}_k$ and $\lambda_k$
(since the first-order mean values $\left\langle \widehat{\mathrm{q}}_{\mu }\right\rangle$
are equal to zero in the case under study):
\be
\mathfrak{\bar{n}}_{m} = \frac12\Big(\sum_{k=1}^4 \widetilde{\lambda}_k \lambda_k\Big)^{(mm)} -\frac12.
\label{Nmlam}
\ee
These mean numbers do not depend on the ``fast time'' $t$. Their explicit expressions are as follows:
\begin{align}
\mathfrak{\bar{n}}_{m} & =\frac{1}{4\eta ^{2}}\Big\{
\left[ \eta+1 +\left(1+\mathfrak{u}_{m}\right)\beta^{2}\left(\eta +\beta^{2}\right)\right]\mathbb{S}_{\mu }^{2}
\notag \\
& + \left[ \eta-1 +\left(1+\mathfrak{u}_{m}\right)\beta^{2}\left(\eta -\beta^{2}\right)\right]\mathtt{S}_{\nu }^{2}
\notag \\
& -2\beta^2\left(\mathbb{C}_{\mu }\mathtt{C}_{\nu } -1\right)
+2\beta \mathfrak{u}_{m} \mathbb{S}_{\mu }\mathtt{S}_{\nu }
\Big\},
\label{gpmn}
\end{align}
where $\mathfrak{u}_{1}=1$ and $\mathfrak{u}_{2}=-1$.
We have introduced the short notation
\begin{align}
\mathbb{C}_{\mu} & =\cosh(\tau_{\mu}), \qquad & \mathbb{S}_{\mu} =\sinh(\tau_{\mu}), \\
\mathtt{C}_{\nu} & =\cos(\tau_{\nu}), \qquad & \mathtt{S}_{\nu} =\sin(\tau_{\nu}).
\end{align}
Typical regimes of excitations of the two modes are illustrated in Fig. \ref{lnmpn}, where
we plot the photon mean numbers in the cavity mode and the detector as functions of the ``dimensionless slow time'' $\tau$
for different values of parameter $\beta$. 
Higher values of $\beta$ imply on a more intense photon generation for a given time.

We see that for $\beta=1$ the numbers of excitations in each mode grow exponentially after a short transient time,
but the number of quanta in the detector mode is significantly smaller than in the cavity mode for any fixed value of $\tau$.
For smaller values of $\beta$ we observe some ``beats'' between the two modes. However, these ``beats'' exist for
limited time intervals, since only the first line in formula (\ref{gpmn}) is important
 for $\tau_{\mu} \gg 1$.
Thus we see that asymptotically the number of quanta in the field mode is always bigger than in the detector
(since $1+\mathfrak{u}_{1}=2$ while $1+\mathfrak{u}_{2}=0$). The ratio can be very big if $\beta \gg 1$: in this case
$\mathfrak{\bar{n}}_{1} /\mathfrak{\bar{n}}_{2} \approx 4\beta^2=(\gamma/\kappa)^2$.
On the other hand, $\mathfrak{\bar{n}}_{1} /\mathfrak{\bar{n}}_{2} =3$ if $\beta =1$, and
$\mathfrak{\bar{n}}_{1} \approx \mathfrak{\bar{n}}_{2}$ for $\beta\ll 1$ (in the asymptotical regime $\tau_{\mu} \gg 1$).

For $\beta \ll 1$ we have $\eta \approx 1 -\beta^2/2$, $\nu\approx 2$, and $\mu\approx \beta$, so that
$\tau_{\mu}\approx \frac12\gamma t$ and $\tau_{\nu}\approx 2\kappa t$.
Therefore the numbers of excitations in the both modes coincide (provided the slow time variable $\tau$ is not very small), with corrections
of the order of $\beta$:
\begin{equation}
\mathfrak{\bar{n}}_{m}^{(\beta\ll 1)}=\frac{1}{2}\mathbb{S}_{\mu }^{2}+\frac{\beta}{2}\mathfrak{u}_{m}\mathtt{S}_{\nu }\mathbb{S}_{\mu }+
\mathcal{O}(\beta^{2}).
\label{n1tr}
\end{equation}
Moreover, the main term in (\ref{n1tr}) does not depend on the field-detector coupling coefficient in this case.
Eq. (\ref{n1tr}) is in agreement with the results obtained in \cite{DK96} by using the method  of slowly varying amplitudes (although
the argument of the $\sinh$ function in \cite{DK96} was twice bigger than here due to a misprint in the definition of parameter
$\mu$ in that paper).

For $\beta=1$ we have $\eta=1$, $\mu =\nu =\sqrt{2}$, and $\tau_{1}=\sqrt{2}\tau $. In this case
\begin{equation}
\mathfrak{\bar{n}}_{m}^{(\beta= 1)} =\frac{1}{2}[\mathbb{S}_{1}^2+\mathbb{C}_{1}(\mathbb{C}_{1}-\mathtt{C}_{1})+\mathfrak{u}_{m}\mathbb{S}_{1}(\mathbb{S}_{1}+\mathtt{S}_{1})] ,
\end{equation}
so that $\mathfrak{\bar{n}}_{1}^{(\beta= 1)}\approx 3\mathfrak{\bar{n}}_{2}^{(\beta= 1)}$ if $\tau_1 \gg 1$.

If $\beta\gg 1$, then $\eta \approx \beta^2 -\frac12$, $\mu\approx 2\beta$, and $\nu \approx 1$, so that
$\tau_{\mu}\approx \gamma t$ and $\tau_{\nu}\approx \kappa t$. Therefore
\begin{align}
\mathfrak{\bar{n}}_{1}^{(\beta\gg 1)} &= \mathbb{S}_{\mu }^{2} + \emph{O}(\beta^{-2}), \\
\mathfrak{\bar{n}}_{2}^{(\beta\gg 1)}& = \frac1{4\beta^2}\left[
\left(\mathbb{C}_{\mu } -\mathtt{C}_{\nu }\right)^2 + 2 \mathtt{S}_{\nu }^{2}\right] + \emph{O}(\beta^{-3}).
\label{n-bigbeta}
\end{align}
The number of photons in the field mode coincides with the result \cite{D95,DK96} obtained in the absence of any detector, up to
corrections of the order of $\emph{O}(\beta^{-2})$.
The number of excitations in the detector is about $\beta^2$ times smaller, and this leads to a significant time delay
in appearance of excitations in the detector: while the photons in the field mode appear after time $t_1 \sim \gamma^{-1}$,
the detector begins to ``feel'' their presence after the time $t_2 \sim t_1 \ln(\beta)$.

This time delay exists also for very small times, since in the limit $\tau\to 0$ Eq. (\ref{gpmn}) can be simplified as follows:
\begin{align}
\mathfrak{\bar{n}}_{1} &= 4\beta^2 \tau^{2} + \emph{O}(\tau^{4}),
\label{n1small}\\
\mathfrak{\bar{n}}_{2} & = \beta^2 \tau^{4} + \emph{O}(\tau^{6}).
\label{n2small}
\end{align}

\section{Photon fluctuations }
\label{sec-fluc}

It is interesting to know, besides the mean numbers of quanta, their variances
$\sigma_n =\langle \hat{n}^2\rangle - \langle \hat{n}\rangle^2$, the degree of squeezing of quadrature
components and the photon distribution functions
 (PDF) ${\cal P}(m)\equiv \langle m|\hat\rho| m\rangle$, where $| m\rangle$ means the $m$th Fock state
 and $\hat\rho$ is the statistical operator describing the mixed (due to the interaction) state of the mode.
For the initial vacuum states of the field mode and detector the time-dependent
statistical operator is Gaussian. Therefore we can use the results obtained in Ref. \cite{VM94} for the most general
Gaussian states.

\subsection{Fluctuations of the photon numbers}

In our case the first-order mean values of the quadrature components are equal to zero, so that
\begin{equation}
\sigma _{\mathfrak{n}_{m}}=2\mathfrak{\bar{n}}_{m}^2 +2\mathfrak{\bar{n}}_{m}-\mathcal{D}_{m} +\frac{1}{4},
\label{sig-n}
\end{equation}
where
\begin{equation}
\mathcal{D}_{m}=\mathcal{M}_{p_{m}p_{m}}\mathcal{M}_{x_{m}x_{m}}-\mathcal{M}_{x_{m}p_{m}}\mathcal{M}_{p_{m}x_{m}}
\end{equation}
is the  \emph{invariant uncertainty product\/} (IUP) of the $m$th mode,
which satisfies the Schr\"{o}dinger--Robertson uncertainty relation \cite{QIBook} $\mathcal{D}_{m}\geq 1/4$
for any quantum state. Besides,
it is easy to verify that
$\mathcal{D} \le \left(\mathfrak{\bar{n}} +1/2\right)^2$
for the states with zero first-order mean values of the quadrature components. Therefore the variances of the photon number in the
Gaussian states with zero first-order mean values of the quadrature components must obey the inequality
$
\sigma _{\mathfrak{n}} \ge \mathfrak{\bar{n}}\left(\mathfrak{\bar{n}} +1\right)$.
%
The equality sign is attained for {\em thermal\/} (i.e., mixed) quantum states. On the other hand, for pure
squeezed vacuum states with $\mathcal{D}=1/4$ we have
\be
\sigma _{\mathfrak{n}}^{(sqzvac)} =2 \mathfrak{\bar{n}}\left(\mathfrak{\bar{n}} +1\right).
\label{sigsqz}
\ee

The calculations lead us to the identical expressions for the functions $\mathcal{D}_m(t)$ for $m=1$ and $m=2$
\begin{align}
\mathcal{D}_{m}& = \frac14 +\frac{1}{8\eta ^{4}}\Big\{
\eta ^{3}\left( \mathbb{S}_{\mu }^{2}+\mathtt{S}_{\nu}^{2}\right)
-4\beta ^{2}\left(\beta^{2}-1\right)\mathbb{S}_{\mu }^{2}\mathtt{S}_{\nu}^{2}
\notag \\
& +\left(3\eta^2 +\eta^2\beta^2 -2\right)\left( \mathbb{S}_{\mu }^{2}-\mathtt{S}_{\nu}^{2}\right)
-2\beta\eta^2 \mathbb{C}_{\mu }\mathbb{S}_{\mu }\mathtt{C}_{\nu }\mathtt{S}_{\nu }
\notag \\
&+4\beta^{3}\left(\mathbb{C}_{\mu }\mathtt{C}_{\nu } -1\right)\left(\mathbb{S}_{\mu }\mathtt{S}_{\nu} -\beta^3\right)
\Big\}.
\label{Dmt}
\end{align}
This result may seem surprising at first glance, since other statistical properties (such as the mean number of quanta
or degree of squeezing  studied in the next subsection) in each mode are  different. But this is the consequence of the property
of IUP for the Gaussian states: for these states IUP determines the quantum purity according to the formula
$
\mbox{Tr}\left(\hat\rho^2\right)= (4\mathcal{D})^{-1/2}$.
On the other hand, it is known that for any {\em pure\/} bipartite quantum state the purities of each part are identical.
Since the initial state of the total system is chosen to be pure (vacuum), it remains pure for any instant of time
in the absence of dissipation (assumed in this paper). Therefore $\mathcal{D}_{1}(t) \equiv \mathcal{D}_{2}(t)$.

Simple formula for variances can be obtained if $\beta \ll 1$:
\begin{equation}
\sigma _{\mathfrak{n}_{m}}^{(\beta\ll 1)} =\frac{1}{4}(2\mathbb{C}_{\mu }^{2}+1)\mathbb{S}_{\mu }^{2}+\frac{\beta}{4}
\mathtt{S}_{\nu }\mathbb{S}_{\mu }
\mathbb{C}_{\mu }(\mathtt{C}_{\nu }+4\mathfrak{u}_{m}\mathbb{C}_{\mu }).
\end{equation}
Comparing it with (\ref{n1tr}), we conclude that quantum states of the both modes are non-thermal,
although they become highly mixed with the course of time, since the purity in this case diminishes as
$\mbox{Tr}\left(\hat\rho^2\right)=[\cosh(\gamma t/2)]^{-1}$.

The limit (\ref{sigsqz}) is achieved approximately for the field mode if $\beta \gg 1$:
\begin{equation}
\sigma _{\mathfrak{n}_{1}}^{(\beta\gg 1)} = 2\mathbb{S}_{\mu }^{2}\mathbb{C}_{\mu }^{2} + \emph{O}(\beta^{-2}).
\end{equation}
The invariant uncertainty product in this case equals
\be
\mathcal{D}_{m}^{(\beta\gg 1)} = \frac14 + \frac1{4\beta^2}\left[\mathbb{C}_{\mu }^{2} +1
-2\mathbb{C}_{\mu }\mathtt{C}_{\nu }\right]\left[1+ \emph{O}(\beta^{-1})\right].
\label{D-bigbeta}
\ee
Comparing Eqs. (\ref{n-bigbeta}) and (\ref{D-bigbeta}), we conclude that
quantum states of the both modes become significantly mixed when
the mean number of excitations in the detector exceeds the unit value.
For $\tau_{\mu}\gg 1$ we have $\mathcal{D}_{m} \sim \exp(2\tau_{\mu})$, whereas
$\mathfrak{\bar{n}}_{m}^2 \sim \exp(4\tau_{\mu})$. Consequently,
if $\mathfrak{\bar{n}}_{m} \gg 1$, then
$\sigma _{\mathfrak{n}_{m}} \approx 2\mathfrak{\bar{n}}_{m}^2$ for $m=1,2$, like in the squeezed
vacuum state, even if the states are highly mixed.

\begin{figure}[hbt]
\par
\begin{center}
\includegraphics[height=3.40truein,width=3.40truein,angle=0]{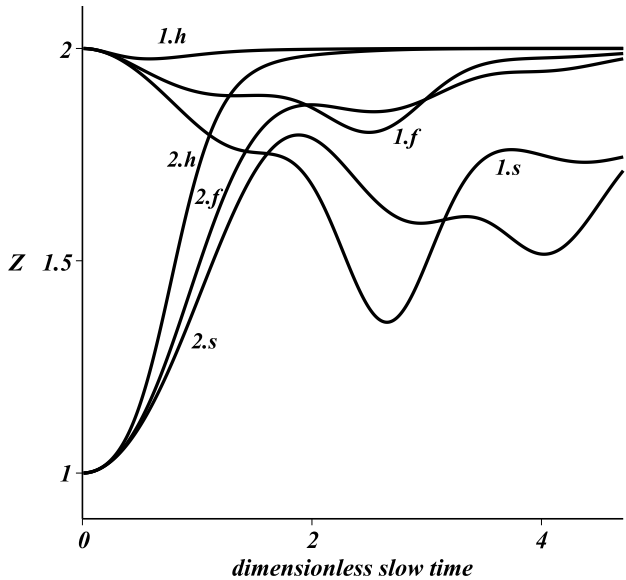}
\end{center}
\par
\vspace{0.1cm}
\caption{The time behavior of
$Z_m \equiv \sigma _{\mathfrak{n}_{m}}/\left[\mathfrak{\bar{n}}_{m}\left(1+\mathfrak{\bar{n}}_{m}\right)\right]$ for the cavity (1) and detector (2) with $\beta=0.2$ (\textit{s}), $\beta=0.5$ (\textit{f}) and $\beta = 1.0$ (\textit{h}).}
\label{fig-rat}
\end{figure}

On the other hand, the photon statistics in the two modes are quite different for $\tau_{\mu}< 1$,
especially in the short-time limit $\tau \to 0$. Frequently the photon statistics is characterized
by the Mandel parameter $Q=\left(\sigma _{\mathfrak{n}} - \mathfrak{\bar{n}}\right)/\mathfrak{\bar{n}}$.
However, in the case concerned this parameter is not very useful, since it is always greater than unity
and grows unlimitedly with time in the DCE regime. A more convenient parameter is the ratio
$Z_m \equiv \sigma _{\mathfrak{n}_{m}}/\left[\mathfrak{\bar{n}}_{m}\left(1+\mathfrak{\bar{n}}_{m}\right)\right]$,
since it varies between $1$ and $2$ for the Gaussian states with zero mean values of quadrature components.
The evolution of functions $Z_1(\tau)$ and $Z_2(\tau)$ for different values of parameter $\beta$
is illustrated in Fig. \ref{fig-rat}.
The difference is clearly seen in the limit $\tau \to 0$,
when Eq. (\ref{Dmt}) can be simplified as follows:
\be
\mathcal{D}_{m}(\tau) = \frac14 +\beta^2 \tau^{4} \Big [ 1+ \frac{2}{9}(3\beta^2-4) \tau^{2} \Big ] + \emph{O}(\tau^{8}).
\label{Dmtsmall}
\ee
Therefore the contribution of $\mathcal{D}_{m}$ can be neglected in Eq. (\ref{sig-n}) with $m=1$,
in view of Eq. (\ref{n1small}), so that
$Z_1(0)=2$. But  $\mathcal{D}_{2}$ has the same order of magnitude as $\mathfrak{\bar{n}}_{2}$ for $\tau\to 0$, according to Eqs. (\ref{n2small}) and (\ref{Dmtsmall}). Therefore $Z_2(0)= 1$. More precisely,
\begin{align}
Z_1(\tau) & =2-\frac{\tau^2}{4} \Big [ 1 - \frac{5}{36}\tau^2 \Big] + \frac{7}{6}\beta^2\tau^4 + \emph{O}(\tau^{6}),
 \nonumber  \\
Z_2(\tau) & =1+\frac{\tau^2}{9} \Big [ 4 - \frac{19}{45}\tau^2 \Big] + \frac{128}{135}\beta^2\tau^4 + \emph{O}(\tau^{6}).
 \nonumber
\end{align}

\subsection{Squeezing}

The degree of squeezing is characterized by the {\em invariant squeezing coefficient\/} (ISC) \cite{JOBrev}
(which is equivalent to the {\em principal squeezing\/} introduced in \cite{ps})
\be
\chi = \frac{4\mathcal{D}}{2\mathfrak{\bar{n}}+1 +\sqrt{(2\mathfrak{\bar{n}}+1)^2 -4\mathcal{D}}}.
\label{ISC}
\ee
This is twice the minimal value of variance of any quadrature component taken over the period
of fast oscillations. For $\tau_{\mu}\gg 1$ we have the asymptotical formula
\begin{align}
\chi_m^{(as)} &= \mathcal{D}/\mathfrak{\bar{n}}_m \notag \\
&= \Big[\eta^2\left(\eta+\beta^2\right) +\left(3\eta^2 -2\right)\mathtt{C}_{\nu }^2
+\left(2-\eta^2\right)\mathtt{S}_{\nu }^2 \notag \\
& -2\beta\left(\eta ^{2}-2\beta^{2}\right)\mathtt{C}_{\nu }\mathtt{S}_{\nu }\Big] \notag \\
&\times
\left\{2\eta^2\left[\beta^2\left(\eta+\beta^2\right)\left(1+\mathfrak{u}_{m}\right) +1+\eta \right]\right\}^{-1}.
\end{align}
In the special case of $\beta \ll 1$ we obtain $\chi_m^{(as)}=1/2$, in agreement with \cite{D95,DK96}.
A very high degree of squeezing can be obtained in the field mode if $\beta\gg 1$: then
$\chi_1^{(as)}=1/(4\beta^4)$. However, there is practically no squeezing in this limit case in the detector
mode: $\chi_2^{(as)}=1 -\mathtt{C}_{\nu }\mathtt{S}_{\nu }/\beta$.
For intermediate values of parameter $\beta$ the ISC does not go asymptotically to some limit value,
but it exhibits slow oscillations in time with the frequency $\nu\kappa$. For example, for $\beta=1$
we obtain
\be
\chi_m^{(as)}\vert_{\beta=1}= \frac{3+ \sin(2\tau_1)}{4(2+u_m)}.
\ee
This function oscillates between the values $1/6$ and $1/3$ for $m=1$ (the field mode). But it is three
times bigger for the detector mode. It is not difficult to calculate the minimal value of the
coefficient $\chi_2^{(as)}$ as function of slow time $\tau_{\nu}$ for arbitrary values of parameter $\beta$:
\be
\chi_2^{(as)}\vert_{min}= \frac12 + \frac{\beta(\beta-1)}{2[1+\eta(\beta)]}.
\ee
Moreover, the minimum of this function with respect to $\beta$ also can be found analytically.
It is achieved for $\beta_{min}=(\sqrt{3}-1)/\sqrt{2}\simeq 0.52$,
being equal to $(1+\sqrt{3})(\sqrt{3}-\sqrt{2})/2\simeq 0.43$.
Consequently, the detector mode cannot be strongly squeezed.

\subsection{Photon distribution functions}

\begin{figure}[hbt]
\par
\begin{center}
\includegraphics[height=3.5truein,width=3.5truein,angle=0]{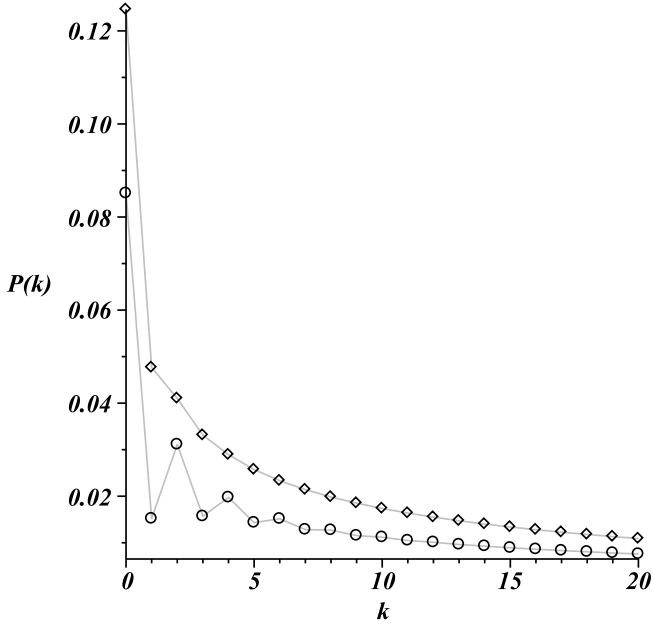}
\end{center}
\par
\vspace{0.1cm}
\caption{The photon distribution functions ${\cal P}_{m}(k)$ for the cavity $(\circ )$ and detector $(\diamond )$
in the case of $\beta = 1.0$ and $\tau = 2$. }
\label{fig-py}
\end{figure}

The PDF of the Gaussian states was derived in \cite{VM94,Chat,Mar}.
For zero mean values of quadrature components $x$ and $p$
it can be expressed in terms of the Legendre polynomials $P_k(x)$ as 
\be
{\cal P}(k)=
\frac{2Y_{-}^{k/2}}{Y_{+}^{(k+1)/2}}
P_k\left(\frac{4\mathcal{D} -1}
{\sqrt{Y_{+}Y_{-}}}\right),
\label{dist0}
\ee
where
\be
Y_{\pm} = 1+4\mathcal{D} \pm 2(1+2\mathfrak{\bar{n}}).
\label{Dpm}
\ee
The behavior of PDF as function of $k$ depends on the value of the argument of the Legendre polynomial.
If this argument is close to zero, then strong oscillations of function ${\cal P}(k)$ are observed, since
Legendre polynomials of zero argument turn into zero for odd values of $k$. Otherwise ${\cal P}(k)$
changes slowly and monotonously. This happens for $\beta\ll 1$, as was shown in \cite{DK96}.
If $\beta\gg 1$, then the argument of the Legendre polynomials for the first mode tends asymptotically
to the small quantity $1/(4i\beta^2)$, so that the PDF shows typical oscillations of the squeezed
vacuum state. But for the second mode the argument of the Legendre polynomials goes asymptotically to the value
$1/(i\sqrt{5})$, which is not small. Therefore the PDF of the detector mode is monotonous function
without oscillations (that agrees with the absence of squeezing in this mode),
which can be well approximated (for $1\ll k \sim \mathfrak{\bar{n}}$)
by the universal dependence derived in \cite{D-PRA09}
\be
{\cal P}(k) \approx \frac{\exp\left[-(2k+1)/(4\mathfrak{\bar{n}})\right] }{\sqrt{\pi\mathfrak{\bar{n}}(2k+1)}}\,.
\ee
In particular, considering formally $k$ as a continuous variable (i.e., using the Euler--MacLaurin summation formula),
we obtain the following value for the probability of detecting any value $k$ smaller than the mean value $\bar{n}$:
\be
\mbox{Prob}(k< \bar{n}) = \mbox{erf}(\sqrt{2}/2)\approx 0.68,
\label{k-erf}
\ee
where $\mbox{erf}(z) \equiv (2/\sqrt{\pi})\int_0 ^z \exp(-x^2)dx$ is the error function.
The distributions ${\cal P}_{1}(k)$ and ${\cal P}_{2}(k)$ for $\beta=1$ and $\tau = 2$ are shown in Fig. \ref{fig-py}.
They are rather different, especially for small values of $k$. The
photon mean numbers are $\mathfrak{\bar{n}}_{1}=112.4$ and $\mathfrak{\bar{n}}_{2}=38.7$.
Nonetheless, the total probabilities to measure the photon number smaller than the $\mathfrak{\bar{n}}_{m}$
are practically the same for each mode: $\mbox{Prob}_1(k< \bar{n})=0.683$ and $\mbox{Prob}_2(k< \bar{n})=0.681$,
in full agreement with Eq. (\ref{k-erf}).

\section{Concluding Remarks}
\label{sec-con}

We have shown that statistical properties of the detector quantum state
(variances of the photon numbers, photon distribution function, and the degree of quadrature squeezing) can be quite
different from that of the field mode. This can be important for planning the experiments on DCE and analysing their results.
In particular, the discovered time delay in the increase of the mean number of quanta in the detector indicates
that influence of losses in the detector (which are not taken into account in the present study) can be essential,
even if losses in the cavity can be neglected. But we leave this problem for the further studies, as well as the
influence of different initial states (e.g., thermal or squeezed) of the field mode and detector on the rate of generation
of the Casimir photons.

\begin{acknowledgments}
The authors acknowledge the partial support of the Brazilian agency CNPq.
\end{acknowledgments}

\appendix

\section{}

Here we give explicit expressions for non-zero coefficients in the solutions (\ref{lambdas}) for
elements of $\lambda_k$-matrices with $k=1$ and $k=3$.

\begin{eqnarray*}
& F_1^{(11)}= \eta+\beta^2,  &\qquad G_1^{(12)}= -\eta+1-\beta(\beta-1), \\
& F_1^{(22)}= \eta-\beta^2,  &\qquad G_1^{(21)}= \eta -1+\beta(\beta+1), \\
& U_1^{(11)}= \eta-\beta^2,  &\qquad V_1^{(12)}= -\eta-1+\beta(\beta-1), \\
& U_1^{(22)}= \eta+\beta^2,  &\qquad V_1^{(21)}= \eta +1-\beta(\beta+1),
\end{eqnarray*}
\begin{eqnarray*}
& f_1^{(12)}= 1+\beta,  &\qquad g_1^{(11)}= \beta+1-(\beta^2+\eta)(2\beta+1), \\
& f_1^{(21)}= -1-\beta,  &\qquad g_1^{(22)}= \beta(\beta+1) +1-\eta, \\
& u_1^{(12)}= -1-\beta,  &\qquad v_1^{(11)}= (\beta^2-\eta)(2\beta+1) - \beta-1, \\
& u_1^{(21)}= 1+\beta,  &\qquad v_1^{(21)}= -\beta(\beta+1) -1-\eta,
\end{eqnarray*}
\begin{eqnarray*}
& F_3^{(12)}= 1-\beta,  &\qquad G_3^{(11)}= 1-\beta+(\beta^2+\eta)(2\beta-1), \\
& F_3^{(21)}= \beta-1,  &\qquad G_3^{(22)}= \beta(\beta-1) +1-\eta, \\
& U_3^{(12)}= \beta-1,  &\qquad V_3^{(11)}= (\eta-\beta^2)(2\beta-1) + \beta-1, \\
& U_3^{(21)}= 1-\beta,  &\qquad V_3^{(22)}= -\beta(\beta-1) -1-\eta,
\end{eqnarray*}
\begin{eqnarray*}
& f_3^{(11)}= -\eta-\beta^2,  &\qquad g_3^{(12)}= \eta-1+\beta(\beta+1), \\
& f_3^{(22)}= \beta^2-\eta,  &\qquad g_3^{(21)}= 1-\eta -\beta(\beta-1), \\
& u_3^{(11)}= \beta^2-\eta,  &\qquad v_3^{(12)}= \eta+1-\beta(\beta+1), \\
& u_3^{(22)}= -\eta-\beta^2,  &\qquad v_3^{(21)}= \beta(\beta-1)-\eta -1.
\end{eqnarray*}
All other coefficients of matrices $\lambda_1$ and $\lambda_3$ are equal to zero.
The coeficients of matrices $\lambda_2$ and $\lambda_4$ are given by
formulas $F_{2,4}^{(ij)} = -f_{1,3}^{(ij)}$, $f_{2,4}^{(ij)} = F_{1,3}^{(ij)}$,
and similar ones for other lower case and capital symbols.

\end{document}